\documentstyle[12pt]{article}
\newcommand{\Int}{\int\limits}

\newcommand{\Sp}[1]{{\mbox{Li}}_2\left(#1\right)}
\begin{document}

\vspace{20mm}

\begin{center}
{\LARGE Standard and hypergeometric \\[3mm]
       representations for loop diagrams \\[3mm]
       and the photon-photon scattering}
\footnote{To appear in Proc. Intern. Conference ``Quarks-92''
        (Zvenigorod, Russia, May 1992);
        World Scientific, Singapore, 1993}  \\[7mm]
{\Large A. I. Davydychev } \\[4mm]
{\large\em Institute for Nuclear Physics,
Moscow State University, \\[2mm]
119899, Moscow, Russia} \\[7mm]
\end{center}

\begin{quotation}
By using the example of photon-photon scattering box diagrams,
different representations for loop integrals are discussed.
A connection between hypergeometric representation and
dilogarithms is derived. An explicit analytic continuation formula to
large values of Mandelstam variables is constructed.
 \end{quotation}

\vspace{0.5cm}

1. When studying many important physical problems (calculation of
radiative corrections to elementary particles interactions,
examination of operator expansions, etc.), we are often confronted
with various types of Feynman loop diagrams. The greatest success
has been achieved in evaluating massless propagator-type diagrams
(depending on one external momentum only). For such diagrams, this
dependence is power-like, and the only problem is how to calculate
numerical factors at these powers. Some powerful methods (Gegenbauer
polynomial technique \cite{GPT}, integration by parts \cite{IBP},
the "uniqueness" method \cite{uniq}, and some others) have been
developed to calculate these numbers, and for some classes of
diagrams the results have been obtained for any numbers of loops
(see, e.g., \cite{multi}).

The problem gets essentially more complicated, however, when we
consider diagrams with masses and (or) with larger number of external
lines. In this cases, dimensionless combinations can be
constructed from squared external momenta and masses. As a result,
we can get complicated functions of these variables (instead of
numbers), and the calculation becomes non-trivial even on one-loop
level. Thus, the problem is how to describe the occurring functions
in all possible regions of momenta and masses. There are different
approaches to this problem.

Commonly-used "standard" representations for loop diagrams usually
involve special functions like polylogarithms ($ \mbox{Li}_N $),
generalized Clausen's functions, etc. Explicit results of such type
are known for one-loop diagrams up to four external lines (see,
e.g., in \cite{'tHV'79,DNS}) and for some particular cases of two-loop
diagrams (see, e.g., \cite{2lQED,UD}). For some more complicated
cases, however, only parametric integral representations are known
(see, e.g., \cite{Kre}).

There exists also another way to represent results for loop diagrams
-- in terms of hypergeometric functions. In the convergence regions,
these functions can be represented as simple power series \cite{Ker}.
Some results of such type have been presented in \cite{AAB+}.
Moreover, in such a way it was possible to obtain explicit formulae
for the general case of one-loop diagrams with arbitrary number
of external lines \cite{JMP}. Note that a special approach
\cite{tens} makes it also possible to represent integrals with
tensor structure in numerators in the same form. If one need to know
results in other regions of variables, it is possible to use
analytic continuation formulae for occurring hypergeometric
functions (see, e.g., in \cite{Bate}). On the other hand, for
complicated cases there is a problem how to construct analytic
continuation to all regions of interest.

To our opinion, both representations can be useful in realistic
calculations. Because of this, it is also important to know
how they are connected (how to go from one representation to
another). Usually it is possible to use parametric integral
representations for hypergeometric functions, and then to try to
evaluate them in terms of known special functions (for example,
polylogarithms). Such a procedure has been already considered
by taking some examples of triangle diagrams \cite{DavDub,INLO2}.
In the present paper we would like to illustrate it via an
example of box diagrams contributing to photon-photon scattering.

\vspace{1cm}

2. Photon-photon scattering ($ \gamma \gamma \rightarrow \gamma
\gamma $) is the well-known example of non-linear effects in QED
(see, e.g., \cite{KN} - \cite{CDeTP} and references therein).
Note that the processes $ g  g \rightarrow \gamma \gamma $ and
$g g \rightarrow g g $ are also described by the same diagrams.
The lowest order of this process contains the box diagram
presented in Fig.~1
\begin{figure}[bth]
\setlength {\unitlength}{1mm}
\begin{picture}(150,60)(0,0)
\put (60,10) {\line(1,0){40}}
\put (60,50) {\line(1,0){40}}
\put (60,10) {\line(0,1){40}}
\put (100,10) {\line(0,1){40}}
\multiput(30,10)(6,0){5}{\line(1,0){3}}
\multiput(30,50)(6,0){5}{\line(1,0){3}}
\multiput(130,10)(-6,0){5}{\line(-1,0){3}}
\multiput(130,50)(-6,0){5}{\line(-1,0){3}}
\put (35,13) {\vector(1,0){12}}
\put (35,47) {\vector(1,0){12}}
\put (113,13) {\vector(1,0){12}}
\put (113,47) {\vector(1,0){12}}
\put (60,10) {\circle*{1}}
\put (60,50) {\circle*{1}}
\put (100,10) {\circle*{1}}
\put (100,50) {\circle*{1}}
\put (22,10) {\makebox(0,0)[bl]{\large $k_1$}}
\put (22,48) {\makebox(0,0)[bl]{\large $k_2$}}
\put (134,48) {\makebox(0,0)[bl]{\large $k_3$}}
\put (134,10) {\makebox(0,0)[bl]{\large $k_4$}}
\end{picture}
\caption{}
\end{figure}
as well as those obtained by permutations ($k_3 \leftrightarrow k_4$)
and ($k_2 \leftrightarrow k_3$). Solid internal lines correspond
to massive fermions (electrons or quarks) with the mass $m$, while
dashed external lines denote massless particles (photons or gluons):
\begin{equation}
\label{massless}
k_1^2 = k_2^2 = k_3^2 = k_4^2 = 0 .
\end{equation}
To simplify the analysis, in what follows we shall restrict ourselves
by scalar integrals corresponding to these box diagrams, because it
is known that integrals with tensor numerators can be reduced to
scalar ones (see, e.g., \cite{PV,tens}). Formally, such scalar
integrals correspond also to massless particles scattering in
the Wick-Cutkosky model.

As usual, we define the Mandelstam variables as
\begin{equation}
\label{stu}
s \equiv (k_1 + k_2)^2 \;  ,  \; \; t \equiv (k_2-k_3)^2 \; ,
\; \; u \equiv (k_1 - k_3)^2 \;  ; \; \; s + t + u = 0 .
\end{equation}
Note that only two of them (for example, $s$ and $t$) are
independent. So, if we denote the integral corresponding to Fig.~1
as $J(s, t; m)$ then the whole amplitude (including diagrams with
permutations) is proportional to the sum
\begin{equation}
\label{permut}
J(s, t; m) + J(s, u; m) + J(u, t; m) .
\end{equation}

Hypergeometric representation for $J$ in four dimensions
can be easily obtained from general formulae of the papers \cite{JMP}
(see also \cite{DeT,Ker}) by putting $k_i^2 = 0$,
\begin{equation}
\label{J}
J(s, t; m) = \frac{i \pi^2}{6 m^4} \;
    F_3 \left( 1, 1, 1, 1; \; 5/2 \left|
         \frac{s}{4 m^2}, \frac{t}{4 m^2} \right. \right) ,
\end{equation}
where
\begin{equation}
\label{F3}
F_3 \left( \alpha, \alpha', \beta, \beta'; \gamma | \; x, y \right)
 \equiv \sum_{j=0}^{\infty}\sum_{l=0}^{\infty}
  \frac{x^j \; y^l}{j! \; l!} \;
  \frac{(\alpha)_j \; (\alpha')_l \; (\beta)_j \; (\beta')_l}
       {(\gamma)_{j+l}}
\end{equation}
is Appell's hypergeometric function of two variables (see, e.g.,
\cite{Bate}), and
\begin{equation}
\label{Poch}
(\alpha)_j \equiv \frac{\Gamma(\alpha + j)}{\Gamma(\alpha)}
\end{equation}
denotes the Pochhammer symbol. Note that the convergence region of
the double series (\ref{F3}) is $ x < 1, \; y < 1 $. So,
formulae (\ref{J})-(\ref{F3}) yield asymptotic expansion of
$J(s, t; m)$ for small values of $x \equiv s/(4 m^2)$ and
$y \equiv t/(4 m^2)$.

To obtain "standard" representation for $J(s, t; m)$, it is
convenient to use the known parametric integral representation
for $F_3$ (see, e.g., \cite{Bate}):
\begin{eqnarray}
\label{intF3}
F_3 \left( \alpha, \alpha', \beta, \beta'; \gamma | \; x, y \right)
= \frac{\Gamma(\gamma)}{\Gamma(\beta) \; \Gamma(\beta') \;
                        \Gamma(\gamma - \beta - \beta')}
\hspace{52mm}
\nonumber \\ [-0.5 cm]
\times \begin{array}{c} {} \\ {} \\
                        \Int_{}^{} \Int_{}^{} \\
                       {}_{ \xi \geq 0, \; \eta \geq 0} \\
                       {}_{ \xi + \eta \leq 1 }
       \end{array}
 d \xi \; d \eta \; \xi^{\beta -1} \; \eta^{\beta' -1} \;
     (1- \xi - \eta)^{\gamma - \beta - \beta' -1} \;
     (1- \xi x)^{-\alpha} \; (1- \eta y)^{-\alpha'}
\hspace{5mm}
\end{eqnarray}
(it is valid for $\mbox{Re} \beta > 0, \; \mbox{Re} \beta' > 0, \;
\mbox{Re}(\gamma - \beta - \beta') > 0$). For the case of (\ref{J})
we find, from (\ref{intF3}), that
\begin{eqnarray}
\label{2intF3}
F_3 \left( 1, 1, 1, 1; \; 5/2 | \; x, y \right)
= \frac{3}{4} \begin{array}{c} {} \\ {} \\
                        \int_{}^{} \int_{}^{} \\
                       {}_{ \xi \geq 0, \; \eta \geq 0} \\
                       {}_{ \xi + \eta \leq 1 }
       \end{array}
 \frac{d \xi \; d \eta}{(1-\xi x)(1-\eta y) \sqrt{1-\xi-\eta}} .
\end{eqnarray}
Evaluating the integrals on the r.h.s. of (\ref{2intF3}), we
arrive at the following result:
\begin{eqnarray}
\label{F3Li2}
F_3 \left( 1, 1, 1, 1; \; 5/2 | \; x, y \right) \hspace{10cm}
\nonumber \\ [0.3 cm]
= \frac{3}{4 x y \beta_{xy}}
 \left\{ 2 \ln^2
    \left( \frac{\beta_{xy}+\beta_x}{\beta_{xy}+\beta_y} \right)
+ \ln \left( \frac{\beta_{xy}-\beta_x}{\beta_{xy}+\beta_x} \right) \;
     \ln \left( \frac{\beta_{xy}-\beta_y}{\beta_{xy}+\beta_y} \right)
   - \frac{\pi^2}{2} \right. \hspace{1cm}
\nonumber \\ [0.3 cm] \left.
  +\sum_{i=x,y}^{} \left[
    2 \Sp{\frac{\beta_{i} -1}{\beta_{xy}+\beta_i}}
   -2 \Sp{-\frac{\beta_{xy}-\beta_i}{\beta_{i} +1}}
  - \ln^2
    \left( \frac{\beta_{i} +1}{\beta_{xy}+\beta_i} \right)
  \right] \right\} ,
\end{eqnarray}
where
\begin{equation}
\label{beta's}
\beta_x \equiv \sqrt{1- \frac{1}{x}}, \; \;
\beta_y \equiv \sqrt{1- \frac{1}{y}}, \; \;
\beta_{xy} \equiv \sqrt{1- \frac{1}{x}- \frac{1}{y}}. \;
\end{equation}
By simple transformations of dilogarithms, it is easy to show that
the result (\ref{J}), (\ref{F3Li2}) coincides with the well-known
expression (see, e.g., \cite{DeT}). Our expression, however, is
more compact and contains four dilogarithms only (instead of
eight ones in \cite{DeT}).

\vspace{1cm}

3. To obtain the formula of analytic continuation to the variables
$1/x$ and $1/y$, it is convenient to use double Mellin--Barnes
representation for $F_3$,
\begin{eqnarray}
\label{MBF3}
F_3 \left( \alpha, \alpha', \beta, \beta'; \gamma | \; x, y \right)
= \frac{\Gamma(\gamma)}{\Gamma(\alpha) \Gamma(\alpha')
                        \Gamma(\beta)  \Gamma(\beta')} \;
   \frac{1}{(2 \pi i)^2}
   \Int_{-i \infty}^{i \infty} \Int_{-i \infty}^{i \infty}
   d a \; d b \; (-x)^a \; (-y)^b \;
\nonumber \\ [0.3 cm]
\times \Gamma(-a) \; \Gamma(-b) \;
\frac{\Gamma(\alpha +a) \; \Gamma(\alpha' +b) \;
             \Gamma(\beta +a) \; \Gamma(\beta' +b)}
            {\Gamma(\gamma + a + b)} , \hspace{0.5cm}
\end{eqnarray}
where the integration contours are chosen so as to separate
"right" and "left" series of poles of gamma functions in the
integrand. If $ |x| < 1, \; |y| < 1 $, we should close both
contours to the right, and we obtain the formula (\ref{F3}).
On the other hand, for large $|x|$ and $|y|$ we should close
the contours to the left. As a result, we obtain the following known
analytic continuation formula (see, e.g., \cite{Bate}):
\begin{eqnarray}
\label{F3F2}
F_3 \left( \alpha, \alpha', \beta, \beta'; \gamma | \; x, y \right)
= \sum_{\{ \lambda, \mu, \rho, \sigma \}}^{}
    \frac{\Gamma(\gamma) \; \Gamma(\rho -\lambda) \;
                            \Gamma(\sigma -\mu)}
         {\Gamma(\rho) \; \Gamma(\sigma) \;
                          \Gamma(\gamma -\lambda -\mu)} \;
   \left( -\frac{1}{x} \right)^{\lambda}
   \left( -\frac{1}{y} \right)^{\mu}
\nonumber \\ [0.3 cm]
\times F_2 \left( \lambda +\mu -\gamma +1, \; \lambda, \; \mu ;
                  \; \lambda -\rho +1, \; \mu -\sigma +1
           \left| \frac{1}{x} , \frac{1}{y} \right. \right) ,
\hspace{0.7cm}
\end{eqnarray}
where $F_2$ is another Appell's hypergeometric function of two
variables,
\begin{equation}
\label{F2}
F_2 \left( \alpha, \beta, \beta'; \; \gamma, \gamma' | \; u, v \right)
 \equiv \sum_{j=0}^{\infty}\sum_{l=0}^{\infty}
  \frac{u^j \; v^l}{j! \; l!} \;
  \frac{(\alpha)_{j+l} \; (\beta)_j \; (\beta')_l}
       {(\gamma)_{j} \; (\gamma')_{l}} ,
\end{equation}
and the sum (in (\ref{F3F2})) extends over the following four sets
of $\{~\lambda,~\mu,~\rho,~\sigma~\}$ :
$\{ \alpha, \alpha', \beta, \beta' \}$,
$\{ \alpha, \beta', \beta, \alpha' \}$,
$\{ \beta, \alpha', \alpha, \beta' \}$ and
$\{ \beta, \beta' , \alpha, \alpha' \}$.
Note that the radius of convergence of the function (\ref{F2}) is
$ u+v < 1$.

The case $ \alpha = \alpha' = \beta = \beta' =1, \; \gamma = 5/2$
corresponds to so-called "logarithmic" case of analytic continuation,
because in this case gamma functions in separate terms on the r.h.s.
of (\ref{F3F2}) have singularities (which cancel in the whole sum).
Considering this limit of the expression
(\ref{F3F2}), we obtain the following analytic continuation formula:
\begin{eqnarray}
\label{F3dF2}
F_3 \left( 1, 1, 1, 1; \; 5/2 | \; x, y \right)
= \frac{3}{4 x y}
 \left\{ \left[ \ln(-4x) \ln(-4y) - \frac{\pi^2}{2} \right]
  F_2 \left( 1/2, 1, 1;  1, 1
          \left| \; \frac{1}{x}, \frac{1}{y} \right. \right)
 \right.
\nonumber \\ [0.3 cm]
- \left[ \ln(-4x) + \ln(-4y) \right] \partial_{\alpha} F_2
- \ln(-4y) \; \partial_{\gamma} F_2- \ln(-4x) \; \partial_{\gamma'} F_2
\nonumber \\ [0.3 cm]
\left. + \partial_{\alpha}^2  F_2
       + \partial_{\alpha} \partial_{\gamma} F_2
       + \partial_{\alpha} \partial_{\gamma'} F_2
       + \partial_{\gamma} \partial_{\gamma'} F_2
 \frac{}{} \right\}, \hspace{2.5cm}
\end{eqnarray}
where we have introduced notation for parametric derivatives of the
function (\ref{F2}), for example:
\begin{equation}
\label{partial}
\partial_{\alpha} F_2 \equiv \frac{\partial}{\partial \alpha}
F_2 \left( \alpha, \beta, \beta'; \; \gamma, \gamma'
  \left| \; \frac{1}{x}, \frac{1}{y} \right)
     \right|_{\alpha =1/2, \; \beta =\beta' =\gamma = \gamma' =1} ,
\end{equation}
and so on. Note that the series representations of these
parametric derivatives (obtained by differentiating the
formula (\ref{F2})) will contain $\psi$-functions ,
\begin{equation}
\nonumber
\psi(z) \equiv \frac{d}{dz} \ln \Gamma(z) \; ,
\end{equation}
and their derivatives, because
\begin{equation}
\label{psi}
\frac{\partial}{\partial \alpha} (\alpha)_j
= (\alpha)_j \; \left( \psi(\alpha + j) - \psi(\alpha) \right),
\end{equation}
etc. The representation (\ref{F3dF2}) gives us asymptotic expansion
of $J(s, t; m)$ (\ref{J}) for large $|x|$ and $|y|$ (with due
regard for $\ln{x}$ and $\ln{y}$ terms).

To be sure that the analytic continuation (\ref{F3dF2}) corresponds
to the same function (\ref{F3Li2}), we may compare the results for
$x \leq 0, \; y \leq 0$. To obtain "standard" representations for
the functions on the r.h.s. of (\ref{F3dF2}), it is convenient
to use the well-known reduction formula,
\begin{equation}
\label{redF2}
F_2 \left( \alpha, \beta, \beta'; \; \beta, \gamma' | \; u, v \right)
= (1-u)^{-\alpha} \;
_2 F_1 \left( \left.
    \begin{array}{c} \alpha, \; \beta' \\ \gamma' \end{array}
       \right|  \frac{v}{1-u} \right) ,
\end{equation}
where $_2 F_1$ is a usual Gauss hypergeometric function. Using the
well-known parametric integral representation for $_2 F_1$,
\begin{equation}
\label{2F1}
_2 F_1 \left( \left.
    \begin{array}{c} \alpha, \; \beta \\ \gamma \end{array}
       \right|  z \right)
= \frac{\Gamma(\gamma)}{\Gamma(\beta) \; \Gamma(\gamma -\beta)} \;
  \Int_0^1 d \xi \; \frac{\xi^{\beta -1} (1-\xi)^{\gamma -\beta -1}}
                       {(1-\xi z)^{\alpha}} \; ,
\end{equation}
and its parametric derivatives, it is easy to obtain the
following results:
\begin{equation}
\label{F2s}
  F_2 \left( 1/2, 1, 1;  1, 1 | \; u, v \right)
   = \frac{1}{\sqrt{1-u-v}} = \frac{1}{\beta_{xy}} \; ,
\end{equation}
\begin{equation}
\label{daF2}
\partial_{\alpha} F_2 = - \frac{2}{\beta_{xy}} \; \ln \beta_{xy} \; ,
\end{equation}
\begin{equation}
\label{dgF2}
\partial_{\gamma} F_2 = - \frac{2}{\beta_{xy}} \;
     \ln \left( \frac{\beta_{xy} +\beta_y}{2 \beta_{xy}} \right) \; ,
\end{equation}
\begin{eqnarray}
\label{dagF2}
\partial_{\alpha} \partial_{\gamma} F_2
  = \frac{2}{\beta_{xy}} \left\{ 2 \ln \beta_{xy} \;
     \ln \left( \frac{\beta_{xy} +\beta_y}{2 \beta_{xy}} \right) \;
  + \Sp{ \frac{\beta_{xy} -\beta_y}{\beta_{xy} +\beta_y}}
  - \Sp{ -\frac{\beta_{xy} -\beta_y}{\beta_{xy} +\beta_y}} \right\}.
\end{eqnarray}
Note that analogous results for $\partial_{\gamma'} F_2$ and
$\partial_{\alpha} \partial_{\gamma'} F_2$ can be obtained from
(\ref{dgF2}) and (\ref{dagF2}) by substitution $x \leftrightarrow y$.

To calculate $\partial_{\gamma} \partial_{\gamma'} F_2$, it is
impossible to use (\ref{redF2}), and we need to employ a double
integral representation for $F_2$ \cite{Bate}. Differentiating it
with respect to $\gamma$ and $\gamma'$ and putting
$\alpha =1/2, \; \beta =\beta' =\gamma = \gamma' =1$, we get
\begin{equation}
\label{intddF2}
\partial_{\gamma} \partial_{\gamma'} F_2
= \frac{3}{4} \; u v \Int_0^1 \Int_0^1  d \xi \; d \eta \;
      \ln(1-\xi) \; \ln(1-\eta) \; (1- u \xi - v \eta)^{-5/2} \; .
\end{equation}
Evaluating the integrals yields
\begin{eqnarray}
\label{dgg'F2}
\partial_{\gamma} \partial_{\gamma'} F_2
= \frac{2}{\beta_{xy}} \; \sum_{i=x,y}^{} \left\{
   \ln^2 \left( \frac{\beta_{xy} + \beta_i}{2 \beta_{xy}} \right)
  - \frac{1}{2}
    \ln^2 \left( \frac{\beta_i +1}{\beta_{xy} + \beta_i} \right)
\hspace{4cm} \right.  \nonumber \\ [0.3 cm]  \left.
     + \! \Sp{\frac{\beta_{i} -1}{\beta_{xy}+\beta_i}}
  \! - \! \Sp{-\frac{\beta_{xy}-\beta_i}{\beta_{i} +1}}
  \! - \! \Sp{ \frac{\beta_{xy} -\beta_i}{\beta_{xy} +\beta_i}}
  \! + \! \Sp{ -\frac{\beta_{xy} -\beta_i}{\beta_{xy} +\beta_i}}
       \right\}.
\end{eqnarray}

Inserting the expressions (\ref{F2s})-(\ref{dagF2}) and
(\ref{dgg'F2}) into (\ref{F3dF2}), we obtain the same result as in
(\ref{F3Li2}). This fact confirms the correctness of analytic
continuation from the variables $x$ and $y$ to $1/x$ and $1/y$.

\vspace{1cm}

4. In the present paper we have discussed different representations
for Feynman loop diagrams by taking, as an example, the box diagram
contributing to the scattering of photons by photons. Hypergeometric
representation (\ref{J}) is convenient for studying asymptotic
expansions and constructing analytic continuation to other variables
(\ref{F3dF2}). Standard representation (\ref{F3Li2}) (involving
dilogarithms of rather complicated arguments) is convenient in
numerical calculations. Anyway, the formula of connection
(\ref{F3Li2}) enables one to pass from one representation to another
(when it is necessary). The formulae (\ref{J}) and
(\ref{F3dF2}) explicitly give all terms of asymptotic expansions for
small and large values of $|s|$ and $|t|$, respectively. Note that
these formulae do not cover all possible regions of Mandelstam
variables (in other regions we need to use other analytic
continuation formulae). For example, in the papers
\cite{Ker,DavDub} analytic continuation to physical thresholds
has been studied. We consider the presented analytic continuation
as an interesting example of using the multiple hypergeometric
functions theory in loop diagrams calculation.

\vspace{0.5cm}

{\bf Acknowledgements}. The author is grateful to E.E.Boos for
drawing his attention to importance of studying $\gamma \gamma$
interactions, and also to R.Scharf for useful discussions.

\end{document}